\input mssymb.tex

\def\inv{^{\raise.15ex\hbox{${
  \scriptscriptstyle -}$}\kern-.05em 1}}

\def\Dsl{\,\raise.15ex\hbox{$/$}\mkern-13.5mu D}
\def\dsl{\raise.15ex\hbox{$/$}\kern-.57em\hbox{$\partial$}}

\def\lspace{\ifx\answ\bigans{}\else\qquad\fi}

\def\CR{\hbox{{$\cal R$}}}

\def\darr#1{\raise1.5ex\hbox{$\leftrightarrow$}
\mkern-16.5mu #1}

\def\INT{{\textstyle \int\kern-.642em\int}}

\def\C{{\Bbb C}}
\def\Z{{\Bbb Z}}

\def\small{\scriptstyle}
\def\tens{\mathop{\otimes}}

\def\la{{\triangleright}}\def\ra{{\triangleleft}}

\def\nquad{{\!\!\!\!\!\!}}

\def\eqn#1#2{\begin{equation}#2\label{#1}\end{equation}}

\def\und#1{{\underline {#1}}}

\def\text#1{\mbox{\rm #1}}
\def\note#1{}

\def\frac#1#2{{{#1\over#2}}}

\def\und#1{{\underline{#1}}}

\def\vect{{\bf t}}\def\vecs{{\bf s}}\def\vecv{{\bf v}}\def\vecu{{\bf
u}}\def\vecx{{\bf x}}

\def\<{\langle}
\def\>{\rangle}

\documentstyle[11pt]{article}
\begin{document} \baselineskip 11pt
\begin{center} {INFINITE BRAIDED TENSOR PRODUCTS AND 2D QUANTUM GRAVITY}
{\ }\\ {\ }\\
{\ }\\ {\small S. MAJID}\footnote{SERC Fellow and Drapers Fellow of
Pembroke College, Cambridge}\\ {\ }\\
{\it Department of Applied Mathematics \& Theoretical Physics}\\ {\it
University of Cambridge, Cambridge CB3 9EW, U.K.}
\end{center}
\begin{quote}
\vskip 5pt
\centerline{\small ABSTRACT} \small
Braided tensor products have been introduced by the author as a
systematic way of making two quantum-group-covariant systems interact
in a covariant way, and used in the theory of braided groups. Here we
study infinite braided tensor products of the quantum plane (or other
constant Zamolodchikov algebra). It turns out that such a structure
precisely describes the exchange algebra in 2D quantum gravity in the
approach of Gervais. We also consider infinite braided tensor products
of quantum groups and braided groups.
\end{quote} \baselineskip 13.7pt

1. Of central importance in the theory of braided groups initiated in
\cite{Ma:bra}\cite{Ma:exa}\cite{Ma:sta}\cite{Ma:lin}\cite{Ma:skl} is
the notion of covariant algebras and their braided tensor products.
Here we give a new application of this notion.

Let us recall that a covariant algebra is simply an algebra $B$ on
which a quantum group $H$ acts (or a dual quantum group $A$ of function
algebra type, coacts) in such a way that the product and unit of $B$
are covariant, i.e. the maps $\cdot : B\tens B\to B$ and $\eta: \C\to
B$ respectively are interwiners for the quantum-group (co)action.  This
is obviously an important notion if we want to work with physical
systems (described by algebras) with  quantum-group symmetry. A fancy
way to think of covariance is that the algebra $B$ {\em lives in} the
category of (co)-modules of the quantum group\cite[Sec. 6]{Ma:exa}:
`Working in the category' just says that we keep everything manifestly
covariant. Thinking about things this way allows us to treat covariance
in the same way as we treat super-symmetry (we make everything
$\Z_2$-graded). This unification between usual notions of group
covariance and super-symmetry is one of the remarkable unifications
made possible by quantum groups\cite[Sec. 6]{Ma:exa}.

In particular, the (co)modules of a true quantum group (with universal
R-matrix or its dual concept for a quantum function algebra) have a
braiding $\Psi$\cite[Sec. 7]{Ma:qua}. This is a coherent collection of
maps $\Psi_{V,W}:V\tens W\to W\tens V$ allowing the `transposition' of
any two objects in the category (any two representations) with
properties like the usual transposition or super-transposition
$\Psi_{V,W}(v\tens w)=(-1)^{\vert v\vert |w|}w\tens v$, except that
typically we no longer have $\Psi_{V,W}=\Psi^{-1}_{W,V}$ (so they are
best represented by braids rather than by permutations).

The main lemma which will concern us, which is the {\em fundamental
lemma of the theory of braided groups} is: if $B,C$ are covariant
algebras then there is a {\em braided tensor product} $B\und\tens C$
which is again a covariant algebra. Thus, we have a way of combining
covariant systems in a covariant way. This braided tensor product
algebra is $B\tens C$ as a space, but product \eqn{e1}{ (b\tens
c)\cdot(d\tens e)=b\Psi_{C,B}(c\tens d)e,\quad b,d\in B,\ c,e\in C}
where we mean to first apply $\Psi_{C,B}$ to $c\tens d\in C\tens B$ and
multiply the result from the left in $B$ and from the right in $C$.
That this is associative follows from functoriality and hexagon
identities for $\Psi$. See \cite{Ma:sta} for a diagrammatic proof. The
braided tensor product $B\und\tens C$ contains $B$ and $C$ as
subalgebras, but unlike the usual tensor product, these subalgebras do
not commute. Instead, their exchange is given by the braiding $\Psi$.

This braided tensor product is like the super-tensor product of
super-algebras, leading us to interpret (\ref{e1}) as saying that the
elements in the algebras $B,C$ have braid statistics rather than
bose-fermi statistics. A braided group then is a generalized Hopf
algebra with $\und\Delta:B\to B\und\tens B$ an intertwiner (covariant)
and an algebra homomorphism provided the elements are treated with
braid statistics as in (\ref{e1}).

2. Let $B=\Bbb C_q$ (the quantum plane) with generators $x_i$. It is
well-known that this is a covariant algebra for $A=SL_q(2)$ with
generators $t^i{}_j$. In this context covariance is equivalent to
requiring that the coaction $x_i\mapsto x_j t^j{}_i$ (the tensor
product on the right is suppressed) is an algebra homomorphism, i.e.
that $x'_i=x_jt^j{}_i$ obey the relations of the quantum plane also.
The same applies for any $R$-matrix with the algebra $B=V^*(R)$ (the
Zamolodchikov algebra of covector type) given by
$\vecx_1\vecx_2=\vecx_2\vecx_1\lambda R$ and $A$ a quantum group
obtained from the FRT bialgebra $A(R)$. This was proven already in
\cite[Sec. 6]{Ma:qua}.

We now compute the infinite braided tensor product of such an algebra
$B$ with itself. Infact, we compute the $N$-fold product and leave the
reader to take the $N\to\infty$ limit. The generators of
$B^{\und\tens^N}$ are $\{x_i(m)\}_{m=1}^{m=N}$ say, where
$x_i(m)=1\tens \cdots\tens x_i\tens \cdots 1$ (in the $m$'th
position).  Since the element $1$ is bosonic (has trivial braiding with
everything else) we have from Section~1, \eqn{e2}{x_i(m)\cdot
x_j(n)=\cases{1\tens\cdots\tens x_i\tens \cdots\tens x_j\tens\cdots
\tens1& \cr 1\tens \cdots\tens x_ix_j\tens\cdots\tens 1&\cr
1\tens\cdots\tens \Psi(x_i\tens \cdots\tens x_j)\tens\cdots\tens 1
&}\nquad=\cases{x_i(m)x_j(n)& if\ $m<n$\cr x_i(m)x_j(n)&  if\ $m=n$\cr
x_l(n)x_k(m)R^k{}_i{}^l{}_j& if $m>n$ }} Here the braiding is
$\Psi(x_i\tens x_j)=x_l\tens x_k\CR(t^k{}_i\tens t^l{}_j)=x_l\tens x_k
R^k{}_i{}^l{}_j$ as explained in detail in \cite{Ma:lin}. The
expression on the right is in terms of the usual (unbraided) tensor
product, relative to which the braided one is a kind of `ordered
product' of the $x_i$. Also, following a trick in \cite{Ma:seq}, let us
write \eqn{e3}{ R(n-m)=\cases{R& if\ $m>n$\cr \lambda R& if\ $m=n$\cr
R^{-1}_{21}& if\ $m<n$}.} It obeys the parametrized Yang-Baxter
equations \eqn{e4}{
R_{12}(m-n)R_{13}(m)R_{23}(n)=R_{23}(n)R_{13}(m)R_{12}(m-n)} with
discrete spectral parameter\cite{Ma:seq}. Then from (\ref{e2}) we can
write $B^{\und\tens^N}$ as generated by the $\{x_i(m)\}$ with relations
\eqn{e5}{\vecx_1(m)\vecx_2(n)=\vecx_2(n)\vecx_1(m)R(n-m)} in an obvious
compact notation. Because the braided tensor product preserves
covariance, we conclude that $B^{\und\tens^N}$ and its continuous limit
are also covariant.  The coaction is the pointwise one
$x_{i}(m)\cdots\mapsto x_j(m)t^j{}_i$ on each generator, extended
multplicatively.

This is exactly a discretised version of the kind of exchange algebra
that arises in 2D quantum gravity\cite{CreGer}. There, one finds fields
$\xi_i(\sigma)$ where $\sigma\in [0,\pi]\subset S^1$ is a  continuous
parameter and $R(\sigma-\sigma')=\cases{R& if\ $\sigma<\sigma'$ \cr
R^{-1}_{21}& if\ $\sigma>\sigma'$}$. Here $R$ is the usual $SL_q(2)$
matrix and the coaction appears equivalently as an action of
$U_q(sl_2)$ by dualizing.  \note{\eqn{e7}{l^+{}^i{}_j\la x_l=x_k
Rk{}_l{}^i{}_j,\quad l^-{}^i{}_j\la x_l=x_k R^{-1}{}^i{}_j{}^k{}_l}
where $l^\pm$ are the generators in FRT form and acting as a quantum
group on products of the generators.} The same applies for $U_q(g)$.
Note that in this setting $\sigma$ is a spatial world-sheet co-ordinate
and our infinite braided tensor product gives a discretised version of
this. Our interpretation of the physical fields then is that they are
sections of a vector bundle over $[0,\pi]$ except that the copies of
the quantum plane over each point are `braided-independent' in the
sense that they are tensor-producted with braid statistics. In this
sense they are classical objects with the quantum non-commutativity
being interpreted as braid-statistical non-commutativity.

As well as covector type Zamolodchikov algebras we can just as well
have quantum vector algebras $V(R)$ as explained in detail in
\cite{Ma:lin}. The generators are $\{v^i\}$ with relations
$\vecv_1\vecv_2=\lambda R_{12}\vecv_2\vecv_1$. One may take their
infinite braided tensor product just as above and interpret the
continuous limit as some other fields transforming as vectors under
$SL_q(2)$. In the case of the quantum plane, the vector and covector
algebras are $\C_q$ and $\C_{q^{-1}}$ respectively in our
approach\cite{Ma:lin} (for a fixed choice of $\lambda$).

3. Another covariant algebra is the braided group $B(R)$ associated to
any bi-invertible $R$-matrix\cite{Ma:exa}. In compact matrix form the
relations and covariance in \cite{Ma:exa} can be written as
\eqn{e8}{R_{21}\vecu_1 R_{12}\vecu_2 = \vecu_2 R_{21} \vecu_1 R_{12},
\quad \vecu\mapsto \vect^{-1}\vecu\vect.} Here $u^i{}_j$ are the
generators. Note that the relations here are {\em not} the reflection
equations arising in lattice models with boundary\cite{Skl:bou} (these
have additional transpositions even in the constant case).  In fact,
they arise in the theory of braided groups by a systematic process of
transmutation of the FRT bialgebras $A(R)$ (hence the name). They have
also been used previously in a compact description of $U_q(g)$ in FRT
form, for reasons explained in \cite{Ma:skl}.

Setting $B=B(R)$ we can compute $B^{\und\tens^N}$ in a similar way to
that above. Thus $B(R)^{\und\tens^N}$ has generators $u^i{}_j(m)$ with
\eqn{e9}{\vecu_1(m)\cdot \vecu_2(n)=\cases{\vecu_1(m)\vecu_2(n)&
if\ $m\le n$\cr \Psi(\vecu_1(m)\tens\vecu_2(n))& if $m>n$ }.} The
braiding between $\vecu$ and itself is computed in \cite{Ma:exa}. In
compact form it can be written as\cite{Ma:lin}
$\Psi(R^{-1}_{12}\vecu_1 \tens R_{12}\vecu_2)=\vecu_2 R^{-1}_{12}\tens
\vecu_1 R_{12}$ so this is relevant when $m>n$. Comparing this with the
relations (\ref{e8}) pertaining when $m=n$, we see that
$B(R)^{\und\tens^N}$ is defined by the relations \eqn{e10}{
R_{21}(m-n)\vecu_1(m) R_{12}(n-m)\vecu_2(n) = \vecu_2(n) R_{21}(m-n)
\vecu_1(m) R_{12}(n-m)} with $R(n-m)$ as in (\ref{e3}) (with
$\lambda=1$). It has a braided-coproduct $\und\Delta
\vecu(m)=\vecu(m)\und\tens\vecu(m)$.

Once again, by construction we will preserve a global symmetry of these
equations under the conjugation (quantum adjoint) coaction
$\vecu(m)\mapsto \vect^{-1}\vecu(m)\vect$ at each site. When $R$ is the
$SL_q(2)$ matrix then $B(R)$ is the degenerate Sklyanin
algebra\cite{Ma:skl} (after setting $BDET(\vecu)=1$ one obtains the
algebra of $U_q(sl_2)$). Assuming both covector fields $\vecx(m)$ and
vector fields $\vecv(m)$ are realized (in the continuum limit) in 2D
quantum gravity, it is interesting to ask if this field of degenerate
Sklyanin algebras is also realised. The reason to believe this is that
locally $u^i{}_j=v^i\und\tens x_j$ in the braided tensor product
$V(R)\und\tens V^*(R)$ is a realization of $B(R)$ \cite{Ma:lin} (these
are the rank-one braided matrices). Since $BDET(\vecu)=0$ in this
realization\cite{Ma:lin}, this would {\em not} be a realization among
the fields of $U_q(sl_2)$ but only of the degenerate Sklyanin algebra.
Nevertheless, this object is a group-like object (it is a bialgebra
with braid statistics, with the matrix coproduct). We have argued in
\cite{Ma:bra} that it is just such group-like objects with braid
statistics (not necessarily quantum groups) that should be realised as
{\em internal} symmetries, i.e. among the fields of a general
low-dimensional quantum theory. Details will be presented elsewhere.

4. Next, let us consider $B=A(R)$ itself as a covariant algebra under
the coaction $\vecs\mapsto \vecs\vect$ where $\vecs$ denotes the
generators of $B$ (as distinguished from the generators $\vect$ of the
coacting quantum group $A$ obtained from $A(R)$. Clearly,
$A(R)\und\tens A(R)$ consists of two copies of $A(R)$ with
cross-relations given by
\eqn{e11}{(1\tens \vecs_1)\cdot(\vecs_2\tens 1)=\Psi(\vecs_1\tens\vecs_2)
=\vecs_2\tens\vecs_1\CR(\vect_1\tens\vect_2)=\vecs_2\tens\vecs_1 R_{12}
=(\vecs_2\tens 1)\cdot(1\tens\vecs_1)R_{12}.}
Similarly in general the algebra $A(R)^{\und\tens^N}$ has generators
$\vecs(m)$ and relations
\eqn{e13}{\vecs_1(m)\vecs_2(n)=\cases{\vecs_2(n)\vecs_1(m) R^{-1}_{21}
&if\ $m<n$\cr R_{12}^{-1}\vecs_2(n)\vecs_1(m)R_{12}& if\ $m=n$\cr
\vecs_2(n)\vecs_1(m)R_{12}& if\ $m>n$}.}
We note here a striking similarity with the exchange algebra among the
fields $u_i,u(x)$ in~\cite{AFS:hid} in their approach to the
discretisation of Kac-Moody algebras. To coincide precisely
we would need $\vecs_1\vecs_2=R_{21}^{-1}\vecs_2\vecs_1 R_{12}$ at $m=n$,
in contrast to (\ref{e13}). In our case, we preserve the quantum-group
symmetry in the form $\vecs(m)\mapsto \vecs(m)\vect$.

5. For completeness, let us mention a previous iterated construction,
not a braided tensor product but a {\em double cross product} of
ordinary Hopf algebras $H_1\bowtie H_2$. This was introduced in
\cite[Sec. 3.2]{Ma:phy} and can be thought of as the quantum analogue of a
Manin
triple (it factorises into $H_1,H_2$). It was explicitly constructed as
a double semidirect product where each factor acts on the other (hence
the notation). We showed that Drinfeld's quantum double is an example
by the mutual quantum coadjoint actions\cite[Sec. 4]{Ma:phy}.

Moreover, as quite a different example from the quantum double, we
showed in \cite{Ma:seq} that $A(R)$ indeed acts on itself by quantum
adjoint actions, so we can form $A(R)\bowtie A(R)$. The actions between
the two copies of $A(R)$ with generators $\vecs,\vect$ are\cite[Thm.
3.2]{Ma:seq}
\eqn{e14}{ \vecs_1\la\vect_2=R_{12}^{-1}\vect_2 R_{12}, \qquad
\vecs_1\ra\vect_2=R^{-1}_{12}\vecs_1R_{12}} and the resulting structure
came out as two copies of $A(R)$ with cross relations \eqn{e15}{
R_{12}\vecs_1\vect_2=\vect_2\vecs_1R_{12}.} Recently these double cross
products of $A(R)$ have attracted attention as some kind of
`complexification' of $A(R)$. We would like to point out that in
\cite{Ma:seq} we went further and iterated this construction. The
infinite double cross product $A(R)^{\bowtie^N}$ comes out\cite[Prop.
3.4]{Ma:seq}
as generated by $\{\vect(m)\}$ with relations\eqn{e16}{
R_{12}(m-n)\vect_1(m)\vect_2(n)=\vect_2(n)\vect_1(m)R_{12}(m-n).} This
is a discrete version of the kind of bialgebra that is useful in
solving lattice models. In contrast to the above iterated braided
tensor products, it does not preserve a quantum action of $A(R)$.

\end{document}